\documentclass[12pt]{article}
\usepackage{graphicx}
\usepackage{hyperref}
\title{Visualising Matter and Cosmologies: \\ A Transhistorical Example\footnote{Published in {\it Column 7 New Imaging: Transdisciplinary strategies for art beyond the new media}, ed. by Su Baker, Melanie Oliver and Paul Thomas (Sydney: Artspace Visual Arts Centre Ltd, 2011, ISSN 1835-3487).\newline \url{http://www.artspace.org.au/publications_column.php?i=7}}}
\author{Luc\'{\i}a Ayala\footnote{Humboldt University of Berlin. E-mail: lucia@luayas.net} \ \&  Jaime E. Forero-Romero\footnote{Astrophysical Institute Potsdam, Astronomy Department UC Berkeley. E-mail: forero@berkeley.edu}}
\date{January 2011}
\begin{document}
\maketitle
\begin{abstract}
We propose a connection between the visualisation of cosmic matter and structure formation in the Cartesian
tradition and that used by contemporary astrophysics. More precisely, we identify cosmological simulations of
large scale structure in the Universe with the system of vortices in Descartes’ physics. This connection operates
at different levels of the images: their representational purpose; the theoretical systems behind their use; and,
finally, their function and materiality as visual productions. A skilled use of image analysis is necessary to stress
the continuities and peculiarities between different epochs and disciplines.
\end{abstract}

\begin{quotation}
{\it La gran lecci\'on filos\'ofica de la ciencia contempor\'anea consiste, precisamente, en
habernos mostrado que las preguntas que la filosof\'{\i}a ha cesado de hacerse desde hace
dos siglos – las preguntas sobre el origen y el fin – son las que de verdad
cuentan\\(Paz 2009: 179).}
\end{quotation}

\newpage

In 1644, the Dutch artist and mathematician Frans van Schooten the Younger (1615-1660) visualised
the system of vortices described by Descartes in his {\it Principia Philosophi\ae} (Fig. 1). Almost 350 years
later, the astrophysicists Melott and Shandarin published the results of their simulations to visualise
large scale cosmological structure in “The Astrophysical Journal” (Fig. 2). These two images appear
very similar, yet for a multitude of reasons they are radically different.

\begin{figure}
\begin{center}
\includegraphics[scale=0.80]{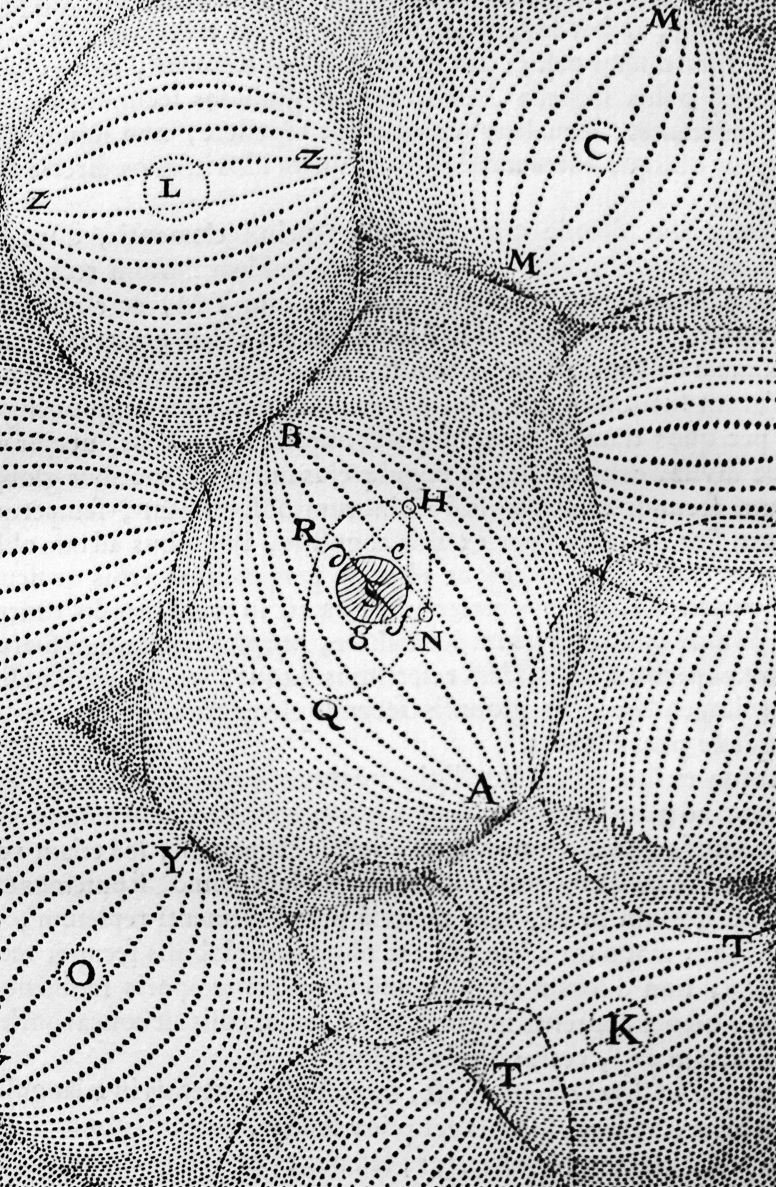}
\end{center}
\caption{Frans van Schooten the Younger, woodcut, system of vortices in Descartes' {\it Principia Philosophi\ae} (Amsterdam: Ludovicum Elzevirium, 1644). Image taken from the 1664 edition. \copyright\ Staatsbibliothek zu Berlin – Preu\ss ischer Kulturbesitz, Abteilung Historische Drucke.}
\end{figure}

\begin{figure}
\begin{center}
\includegraphics[scale=0.40]{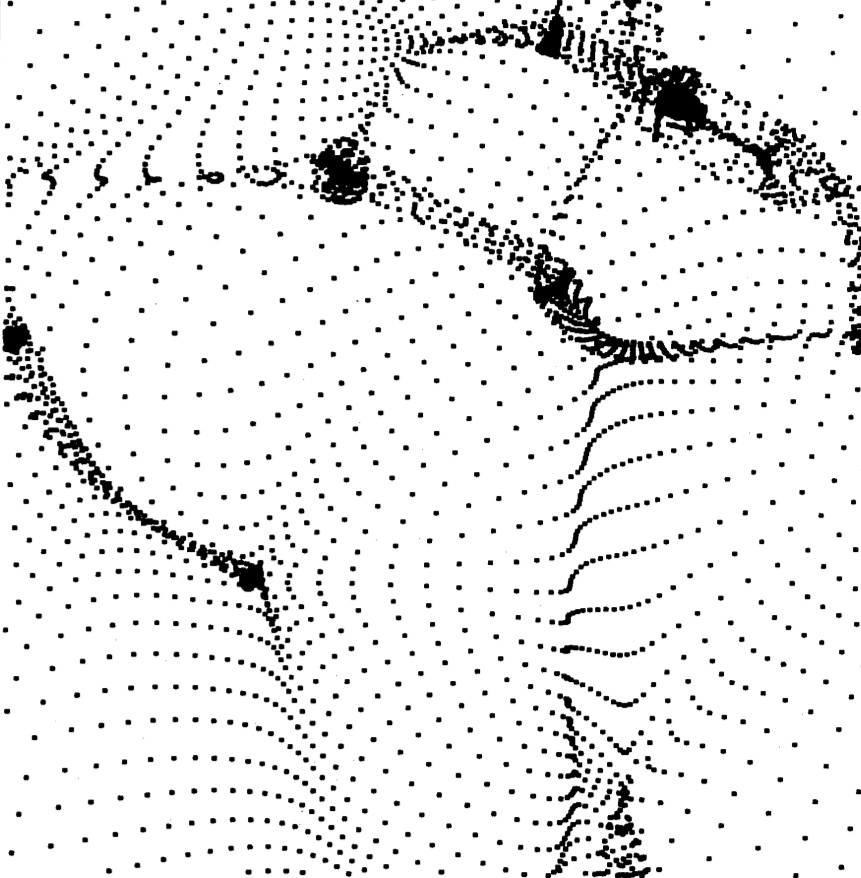}
\end{center}
\caption{Adrian L. Melott and Sergei F. Shandarin, simulation of density perturbations in structure formation on cosmological scales, appeared first in ``Gravitational Instability with High Resolution'', {\it The Astrophysical Journal} 343 (1989): 28. \copyright\ Melott and Shandarin. Reproduction with the kind permission of the authors.}
\end{figure}

\section*{Shattering motion and the Big Bang: Cosmos as history}

Descartes configured the grounds of his physics in {\it Le Monde}, a text written between 1629 and
1633 but withheld from publication until 1664 due to Descartes’ anxiety about the persecution of
Galileo. In {\it Le Monde}, Descartes expounds a theory explaining the formation of the cosmos, but not
its origin, since it was assumed to be a creation of God. Descartes posits an already-created matter
``that should be imagined as the hardest and solidest body existing in the world'' (Descartes 1989
[1664]: 132). Once the initial conditions are set, he describes the dynamics of their evolution to explain
the formation of the actual Universe. At a given moment, God started to shake this compressed
matter in such a way that the shaken parts divided themselves, triggering the motion and subsequent
division of the closest ones in a kind of ``chain reaction''. As a result of this primordial shattering
motion, the matter acquired the most diverse forms, ``like pieces {\it exploding} when a stone is
broken'' (Descartes 1989 [1664]: 136).

Descartes’ model bears remarkable similarities to the contemporary observational framework of
physical cosmology, in which the Universe is understood to have been more homogeneous, denser
and hotter in the distant past than it is today. Like Descartes in {\it Le Monde}\footnote{After the bad fortune of Galileo’s affaire, Descartes postponed the publication of {\it Le Monde} and stressed in his {\it Principia} the role of God in every process. However, when {\it Le Monde} was finally published, the first exposition of his theories was available – consequently we can attest that the relevance of god was not at a first instance the most significant factor.}, contemporary
cosmological theories attempt to explain the {\it formation} of the modern cosmos from specific initial
conditions without seeking their origin. Indeed, a standard theoretical picture of these origins is still
lacking: processes concerning the asymmetry between created matter and anti-matter, and the origin
and nature of the initially small density inhomogeneities that constitute the seeds of present-day
galaxies, for example, are explained by different schemes that await confirmation by future particle
physics experiments.

\section*{Fluid media: subtle matter and dark matter}
In addition to these initial (exploding, expanding) conditions, it is useful to consider several aspects of the properties and implications of the conception of matter deployed by Cartesian and contemporary models, since these notions influence their understanding and explanation of structure formation. A dialogue exists between the Cartesian and contemporary models at the level of their visual components. Moreover, paying attention to the similarities and the differences of the theoretical frames underlying (and manifest through) these visualisations can be extremely productive.

Descartes uses the concept of subtle matter, a fluid composed of particles in constant motion. Van
Schooten depicts its penetrable nature, subjected to constant change and interactions, by dotted
surfaces (see fig. 1)\footnote{For a comparable use of dotted surfaces, see Ayala, Luc\'{\i}a, ``Surpassing human nature: Reinventions of and for the body as a consequence of astronomical experiments in the seventeenth and eighteenth centuries'', {\it Metaverse Creativity} 1: 1 (2010), specially pp. 109-110.}. From a historical perspective, one of the main contributions of his model was
the introduction of the visualisation of matter as a key element in astronomical work. In doing so,
special consideration was given to the {\it quality of matter as essence of the model}. When other authors
popularized the Cartesian model, in most cases only this dotted surface was highlighted. The
visualisation of cosmic matter and how it forms the general structure, started to be more important
than concrete phenomena or mathematical laws. This became a distinctive visual feature of Cartesian
physics, even in abstract figures. As a consequence, the traditional way of presenting astronomical diagrams (considering only orbital trajectories and the position/organisation of bodies) was enriched by the depiction of cosmic matter. An inspection of diagrams depicting the Tychonic and the Cartesian systems in a 1761 English translation of the {\it Conversations on the Plurality of Worlds} by Fontenelle provides an illuminating example\footnote{First published in Paris: Veuve C. Blageart, 1686.} in addition to the concentric circles delineating the orbits, there are dots filling the gaps between them. But this is merely one example within a larger tradition. From 1664 onwards, it was not possible any more to construct an astronomical diagram
without attracting attention to matter as one of its main elements, since matter was an essential factor
in the formation of structures.

This logic of astronomical representation remains with us today. The first physically accurate
representations of large-scale structure formation models, obtained through computational
experiments, also highlight the role of matter visualisations. In scientific papers, the matter distribution
is depicted with dots that represent computational, non-physical, particles. This discretization of
matter into particles is necessary to perform the calculations, even when the physical model
corresponds to a fluid. First implemented in supercomputer simulations in the 1970s, particle-based
models remain a workhorse of computational astrophysics. However, the graphical representation of
these models has evolved significantly, such that new visualisation software renders the simulation
output and yields a closer graphical approximation of the postulated fluid nature of the matter being
simulated. In the images from the Millennium Simulation performed by the Virgo Consortium at the
Max Planck Institute for Astrophysics in 2005, dots spread out into different shades and hues, with
colour palettes evocative of the imagined nature of dark matter. By comparing this kind of image with
the versions in the seminal papers of the field, where the particles are clearly depicted as dots, it is
evident that a higher degree of sophistication in the visual language has enabled the visualisation of
more information. If the Millennium Simulation were represented by dots, all the nuances in the
filaments, voids and knots would disappear, being reduced to a black smudge. Nevertheless, the
basic topological information of the cosmic web would be contained in both cases.

While Cartesian physics is based on ``subtle matter'' composed of particles in motion, contemporary
physical cosmology holds that the dominant matter component in our Universe is ``dark matter''.
However, dark matter is not immediatly compatible with the current framework of particle physics. It emerges as a consequence of our conceptual understanding of gravitation: the equations describing its behaviour
correspond to a collisionless fluid that only interacts through gravity. In other words, currently the
behaviour of dark matter cannot be described from basic principles of particle physics (a dark matter
particle has not been detected yet) and, therefore, the only available approach is through its
gravitational effects. To fully explore these effects, numerical simulations are required.

\section*{Structure formation: haloes and vortices}

We have mentioned the differences between the renderings of the first large-scale structure
simulations and the most recent ones. An important consequence of the increasingly sophisticated
visual language used by this field of research has been the possibility to observe the emergence of
new structures inside the cosmic web, namely high concentrations of dark matter with shapes close
to spherical and having a spinning motion. These concentrations are called {\it haloes}, and they play a
fundamental role in galaxy formation models.

In galaxy formation models, each galaxy is placed inside a dark matter halo that is gravitationally
attracted to other haloes, which can therefore collide and merge. The galaxies inside the haloes can
also fuse, transforming their morphology: a larger galaxy is formed out of two smaller ones. For example, the Milky Way and Andromeda, our closest disk galaxy companion, are expected to merge in five billion years. The resulting shape is expected to be spherical, instead of a disk, as the galactic structure
changes during the merger. This is the basis of the hierarchical picture of galaxy formation, where
structures grow from the merging of smaller ones, while their host dark matter haloes trace the
cosmic web.

This modern chronicle once again resonates with Cartesian physics. According to Descartes’ model,
matter is composed of particles in motion revolving around several centres. This behaviour forms
different systems or vortices, each one described as ``a heaven that spins round the star'' (Descartes
1989 [1664]: 140). The vortices are also labelled as ``large heavens'', being ``very unequal in size''
among each other (Descartes 1989 [1664]: 226). Since they are liquid, the shape of vortices is
supposed to be oval (Descartes 1989 [1664]: 186). The heavenly bodies are placed in the middle of
the vortex to which they belong. This interplay between heavenly bodies and vortices (or ``large
heavens'') recalls galaxies and haloes in contemporary physical cosmology.

Developing this point further, the dynamics of vortices also mirror the hierarchical merging of haloes.
By definition, the Cartesian particles are constantly moving and they may collide, leading to erosion or
fusion. In either case, this collision changes the structure of matter (Descartes 1989 [1664]: 146).
Vortices disappear and their respective centres (the heavenly bodies) approach each other, forming
new structures. Each satellite of Jupiter, for example, was considered by Descartes to be the remains
of an ancient, more complex system, whose original structure was lost due to a collision. When their respective contexts vanished, the satellites moved towards the nearest body (in this case the planet)
and were integrated into the system, forming a new vortex. One of the engravings in the {\it Natural
History} by Buffon, published in 1752, presents God creating the Solar System as a series of fluid
vortices. Each planet is depicted as originally belonging to a separate structure, prior to the present-
day organisation of their orbits around the Sun.

\section*{Images and simulations shaping large-scale\\ structures}

Frans van Schooten was entrusted with the task of visualising the system of vortices theorised by
Descartes; together, both Descartes and van Schooten gave {\it tourbillons} their shape. In 1989, Melott
and Shandarin published a series of technical papers dealing with the simulation and visualisation of
large-scale structure. This essay has suggested that a comparison between the two images can be
undertaken on multiple levels, finding similarities at the representational level and between their
respective theoretical frameworks. To complete this comparison, we will examine some aspects of the
materiality of the images themselves.

Both models present a fragment as synecdoche of the whole: the systems cover the entire surface,
extending themselves to the borders of it. In addition, the structure is composed of repetitive
elements. These two aspects, {\it fragment and repetition}, visually indicate a wider space beyond that
already shown. In other words, what we know about discrete areas of the Universe can be applied to
the whole. Taking into account the limitations of the observations, this factor is quite relevant in order
to achieve a general valid model.
Superposed borders delimit the vortices and interconnect the systems through shared areas,
stressing the penetrability of the fluid medium. Borders do not segregate individualities; instead, they
highlight the fact of belonging to a complex system. In the second example, the visualised
computational particles present filaments connecting large dark matter concentrations.
Descartes emphasized a {\it cluster of vortices}, or penetrable, interconnected, and volumetric entities
containing structures centred around stars, with orbiting planets and satellites; the contemporary
model lays stress on the emergence of a {\it cosmic web}, or a network of large dark matter filaments
interconnecting the most massive galaxy clusters.
For both models, voids are important. Descartes neglected the possibility of a vacuum, due to the fact
that there are particles of matter everywhere in the cosmos. To make this aspect clear, the triangles
originating in the interstices among the systems are covered with dense dotted surfaces. For the dark
matter model, voids (the regions with a sparse particle distribution) are important as well from a
quantitative and technical point of view.

The main divergence between each image lays in their function. On one hand, the vortices are a
{\it visualisation of a theory}, a visual explanation containing the key ideas expounded by Descartes in the
text. In this sense, the engraving is as abstract as the theoretical level itself. On the other hand,
simulations of large-scale structure are a direct result of a concrete need to reproduce the available
observations. They are an {\it indispensable tool to verify the theory}; the success of a model derives from
what is revealed in the simulation. Figure 1 visualises Cartesian theory; Figure 2 visualises the numerical experiment that will shape the theory. For this reason, the images are radically different with
respect to the contexts to which they belong.

The process of producing images has changed radically since the mid twentieth
century. Contemporary scientists have constructed a new relationship to the images that they obtain, generate and analyse. In the specific context that we have discussed, the main improvements have
been achieved through the introduction of simulations. Visually comparing the structures derived from
a simulation and implied by observations plays a vital role in the work of contemporary
astrophysicists; this process of visual inspection and recognition escorts the quantitative labor of
extracting and comparing detailed statistics.

As images, Cartesian vortices and contemporary large-scale structure are quite different from a
technical and functional point of view. But conceptually, again, they are quite similar. In {\it Le Monde},
Descartes presents the whole explanation of his system as a simulation:

\begin{quote}
For a short time, then, allow your thought to wander beyond this world to view another, wholly new one,
which I shall cause to unfold before it in imaginary spaces. (Descartes 1989 [1664]: 99).

And my plan is not to set out (as they – the philosophers – do) the things that are in fact in the true world, but only to make up as I please from [this matter] a [world] in which there is nothing that the densest minds are not capable of conceiving, and which nevertheless could be created exactly the way I have made it up.
(Descartes 1989 [1664]: 107).
\end{quote}

In order to conceive the new physics that he proposes, Descartes commences by establishing the
conditions of a certain configuration of matter; he then applies the laws that he understands as valid
and observes the results in this simulated world. He asserts that the same principles can be applied
to the actual world. The validity of Descartes’ system lies in this comparison or analogy. The same
method of shaping physical theories through simulations is also applied today.

\section*{Conclusions}

A comparable basic understanding of the composition and behaviour of the Universe can be traced
from the Cartesian system of vortices to contemporary physical cosmology. Whether shaped by God
or by the Big Bang, the hard/dense primordial matter started to evolve, forming new structures as
consequence of its expansion. The constitution of cosmic matter – particles in motion within a fluid
medium – is depicted by dotted surfaces in both cases, forming together a long tradition in modern
science in which the visualisation of matter constitutes the basis of the model. Since Descartes,
moreover, theories in physics require simulations to be both explained and shaped.

Our understanding of art history dealing with astronomy does not consist of presenting artistic
projects inspired by this science. On the contrary, we focus our attention on astronomical productions
themselves, that is to say, specific visual materials used by scientists to make science, whether they
are produced by artists, as in the case of Descartes, or by scientists who create their own
visualisations. In any case, what is important to stress here is the necessity of a deep understanding of the image as a crucial component of astronomy. One cannot underestimate its relevance by reducing it to a mere ``representation'' or ``illustration'': scientific images are not illustrations, but \emph{pivotal tools in the process of knowledge production}. The specific case of cosmic matter reveals the important role of visualisations. In the examples outlined above, matter is constructed directly \emph{in and through the image}. Advanced techniques to see the cosmic particles were not at Descartes’ disposal; even today, we do not have a technique to detect dark matter particles directly. In both cases the visualisation of matter is required to make the science evolve. Therefore, the examples we have shown are not representations, because the physical appearance of the subject to be supposedly ``represented'' was and is unknown; but they are the objective results of the knowledge we have attained. From this standpoint, an analysis from a renewed art historical perspective, when applied to science, provides an essential tool for understanding its materials with a new strategy.

\section*{Acknowledgements}
We would like to thank Annie Hughes for proofreading the text.

\section*{References}
\noindent
Descartes, R., {\it Principia Philosophi\ae} (Amsterdam: Ludovicum Elzevirium, 1644).\\
\\
Descartes R., {\it Le Monde. Trait\'e de la lumi\`ere / El Mundo. Tratado de la luz}, bilingual edition, trans. by Salvio Turró (Barcelona: Anthropos, 1989 [1664]).\\
\\
Melott A. L. and Shandarin S. F., ``Gravitational Instability with High Resolution'', {\it The
Astrophysical Journal} 343 (1989): 26-30.\\
\\
Paz  O., ``Rodeos hacia una conclusi\'on'', in {\it La llama doble} (Barcelona: Seix Barral, 2009),
174-203.\\
\\
Springel V., White S. D. M., Jenkins A., Frenk C. S., Yoshida N., Gao L., Navarro J., Thacker R.,
Croton D., Helly J., Peacock J. A., Cole S., Thomas P., Couchman H., Evrard A., Colberg J. and Pearce
F., ``Simulations of the formation, evolution and clustering of galaxies and quasars'', {\it Nature} 435 (2005): 629.

\end{document}